\documentclass[prl,aps,twocolumn,amsmath,amssymb,showpacs]{revtex4}
\usepackage{graphicx}
\begin{document}
\title{Formation and evolution of density singularities in hydrodynamics of inelastic gases}
\author{Itzhak Fouxon}
\author{Baruch Meerson}
\author{Michael Assaf}
\author{Eli Livne}
\affiliation{Racah Institute of Physics, Hebrew University of Jerusalem,
Jerusalem 91904, Israel}
\begin{abstract}
We use ideal hydrodynamics to investigate clustering in a gas of inelastically
colliding spheres. The hydrodynamic equations exhibit a finite-time density
blowup, where the gas pressure remains finite. The density blowup signals
formation of close-packed clusters. The blowup dynamics is universal and
describable by exact analytic solutions continuable beyond the blowup time.
These solutions show that {\it dilute} hydrodynamic equations yield a powerful
effective description of a granular gas flow with close-packed clusters,
described as finite-mass point-like singularities of the density. This
description is similar in spirit to the description of shocks in ordinary ideal
gas dynamics.
\end{abstract}
 \maketitle

A central problem of non-equilibrium physics is to describe clusters of matter
developing from structureless initial states. Examples are numerous and include
the large-scale structure of the Universe \cite{universe}, star formation
\cite{star}, radiation-driven plasma condensations \cite{meersonRMP}, clustering
of inertial particles in turbulent flows \cite{inertial} and clustering in
inelastic gases \cite{Goldhirsch,McNamara} considered in this Rapid
Communication.

Inelastic (or granular) gas, a simple model of granular flow \cite{BP}, is a
dilute assembly of hard spheres which lose energy at collisions. In the simplest
version of the model the normal relative velocity of particles is reduced by a
constant factor $0\leq r<1$ upon each collision. At collisions, internal degrees
of freedom of the particles absorb, apart from the energy, the entropy of the
gas. As a result, inelastic gases exhibit a host of structure forming
instabilities, including the famous clustering instability of a freely cooling
homogeneous inelastic gas
\cite{Goldhirsch,McNamara,Ernst,Brey,Luding,Ben-Naim2,ELM,MP}. This instability
involves the development of macroscopic solenoidal flow (the shear mode) and
potential flow (the clustering mode), the latter causing the formation of
clusters of particles where the particle density ultimately approaches the
density of close packing of spheres \cite{MP}. The physical mechanism behind the
initial density buildup is well understood \cite{Goldhirsch,Field}. Collisional
energy loss, enhanced by a local density excess, can lead to a gas temperature
decrease so strong that the local pressure falls, causing a further gas inflow
and density growth. Pressure, therefore, plays a central role in the initial
density build up, as corroborated by the linear theory of the clustering
instability \cite{Goldhirsch,McNamara}, where the density growth is accompanied
by a decreasing in time pressure, having a local spatial minimum. How do
non-linear effects change this linear scenario? Efrati \textit{et al.}
\cite{ELM} and Meerson and Puglisi \cite{MP} addressed this question in the
geometry of a long and narrow channel which we adopt in this work too. The
channel geometry allows only one-dimensional (1D) macroscopic motions and
therefore eliminates the shear mode. In the limit of strong instability (see
below), Efrati \textit{et al.} \cite{ELM} observed numerically a finite-time
blow up of the gas density in 1D hydrodynamic equations. Is it possible to
establish the density blow up analytically, and does it develop for generic
initial conditions? What is the role of pressure in the density blow up? Can
hydrodynamics of dilute gas describe the state of the gas with already existing
density spikes (interpreted here as close-packed clusters)? These questions are
answered in this Rapid Communication.

As clustering only intensifies with inelasticity, it suffices to consider the
nearly elastic limit: $1-r\ll 1$. This allows one (see Refs.
\cite{BP,Goldhirsch2}) to employ hydrodynamics:
\begin{eqnarray}&&
\partial_t \rho+\partial_x (\rho v)=0,\ \
\rho \left(\partial_t v+v \partial_x v \right) = -\partial_x(\rho
T),\label{massmomcons}\\&& \partial_t T+v \partial_x T=-(\gamma-1)T
\partial_x v-\Lambda \rho T^{3/2}\,, \label{temperature}
\end{eqnarray}
where  $\rho$ is the gas density, $v$ is the velocity, $T$ is the temperature
(the particle mass is put to unity), $p=\rho T$ is the pressure, and $\gamma$ is
the adiabatic index ($\gamma=2$ and $5/3$ in 2D and 3D, respectively). Equations
(\ref{massmomcons}) and (\ref{temperature}) describe a 1D dilute gas flow
\cite{Landau} with bulk energy losses. These are accounted for by the term
$\Lambda \rho T^{3/2}$ coming from $(1-r^2) T$ (proportional to the inelastic
energy loss per collision) times $\rho T^{1/2}$ (proportional to the collision
rate). Kinetic theory yields $\Lambda=2 \pi^{(d-1)/2} (1-r^2) \sigma^{d-1}/[d\,
\Gamma(d/2)]$, where $d>1$ is the space dimension, and $\sigma$ is the particles
diameter, see \textit{e.g.} \cite{Brey}. The omission of viscous and heat
conduction terms in Eqs.~(\ref{massmomcons}) and (\ref{temperature}) is
justified in sufficiently long systems as, during its linear stage, the
clustering instability produces, at $1-r\ll 1$, effective initial conditions
with the Reynolds number $Re \gg 1$ \cite{Goldhirsch,McNamara,viscosity}.

As in other 1D compressible flows, it is convenient to go over from the Eulerian
coordinate $x$ to the Lagrangian mass coordinate $m(x, t)=\int_0^x
\rho(x^{\prime}, t) dx^{\prime}$. Here $m(x, t)$ measures the mass content
between the origin \cite{even} and point $x$. This renders a simpler form of the
equations:
\begin{eqnarray}&&
\partial_t(1/\rho)=\partial_m v,\,\,\, \partial_t v
=-\partial_m p,\nonumber\\&& \partial_t p=-\gamma p \rho\,\partial_m v-\Lambda
p^{3/2}\rho^{1/2}\,. \label{eqs0}
\end{eqnarray}
Importantly, the cooling coefficient $\Lambda$  can be eliminated from
Eqs.~(\ref{eqs0}) owing to the following scaling property: $\rho(m,
t)=\rho_*(\Lambda m/\Lambda_*, \Lambda t/\Lambda_*)$, $v(m, t)=v_*(\Lambda
m/\Lambda_*, \Lambda t/\Lambda_*)$ and $p(m, t)=p_*(\Lambda m/\Lambda_*, \Lambda
t/\Lambda_*)$, where $u_*$, $v_*$ and $p_*$ solve Eqs.~(\ref{eqs0}) for
$\Lambda_*=\gamma\sqrt{2}$. Therefore, in most of the following we put
$\Lambda=\Lambda_*=\gamma\sqrt{2}$.

\begin{figure}%
\includegraphics[width=7.0 cm,clip=]{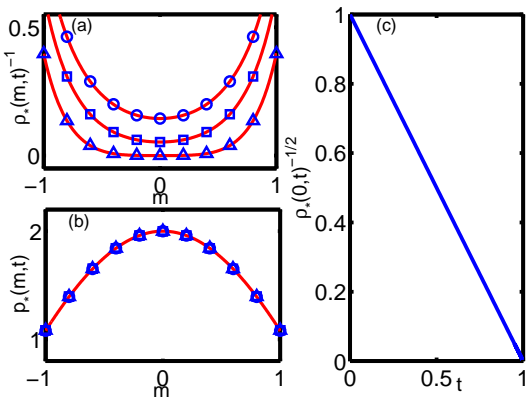}
\caption{The rescaled inverse density (a) and pressure (b) at $t=0.62$
(circles), $0.77$ (squares) and $0.99$ (triangles). The solid lines show the
analytical solution (\ref{particular}). (c): the inverse square root of the
density at the singularity point $m=0$.}\label{A1}
\end{figure}

Motivated by extensive numerical simulations (we used the classic artificial
viscosity, staggered grid scheme of von Neumann and Richtmyer, see Ref.
\cite{code}), we found a family of exact time-dependent analytical solutions of
Eqs.~(\ref{eqs0}), for which an initially smooth density profile blows up in a
finite time $t=t_c$. A distinct feature of these solutions is that the fluid
elements have a constant in time acceleration, so $v(m,t)$ depends on time
linearly, while $p(m,t)$ is time-independent. These solutions are:
\begin{eqnarray}&&
\rho_*(m, t)=\frac{\rho_*(m, 0)}{[1-t\sqrt{A\rho_*(m, 0)\cos m}]^2},\
p_*\!=\!2A\cos m, \nonumber\\&&
\;\;\;\;v_*(m,t)\!=\!-2\!\int_0^m\!\!\sqrt{\frac{A\cos m'}{\rho_*(m',
0)}}dm'\!+\!2A t\sin m\,, \label{special}
\end{eqnarray}
where $\rho_*(m,0)$  is the initial density (an arbitrary function) \cite{even},
and $A>0$ is an arbitrary constant. If $\rho_*(m, 0)$ has a maximum at $m=0$,
then $\rho_*(m , t)$ blows up as $(t_c-t)^{-2}$ at $m=0$ and $t=t_c=[A\rho_*(0,
0)]^{-1/2}$. As $p$ must be non-negative, these solutions hold only at
$|m|\leq\pi/2$, implying a finite mass of the gas $m=\pi$. Depending on the
behavior of $\rho_*(m, 0)$ near $|m|=\pi/2$, this finite mass can be distributed
over either an infinite, or a finite $x$-interval. A simple example of the
former,
\begin{eqnarray}&&
\!\!\!\!\!\!\!\rho_*(m,t)\!=\!\rho_0 \cos m (1-t\sqrt{A \rho_0}\cos m)^{-2},
\;p_*\!=\!2A\cos m,\nonumber\\&& \;\;\;\;\;\;v_*(m,t)=-2 (A/\rho_0)^{1/2}
m+2At\sin m, \label{particular}
\end{eqnarray}
exhibits a density blow up at $m=0$ and $t_c=(A \rho_0)^{-1/2}$. We confirmed
this solution numerically which indicates its stability with respect to small
perturbations, see Fig.~\ref{A1}.

Another exact solution (details will appear elsewhere \cite{long}) shows that an
emerging density blowup may coexist with an ``ordinary" gas dynamic shock, and
the Rankine-Hugoniot conditions at the shock \cite{Landau} are obeyed. This
solution involves, at $t=0$, a zero-temperature gas with density
$\rho_*(m,0)=\rho_0\exp\left(2\sqrt{\gamma-1}\,m\right)\cos^3 m$ on the
Lagrangian interval $0\leq m\leq \pi/2$. The gas moves uniformly, $v_*(x,
0)=-v_0$, and at $t=0$ hits a rigid wall located at $x=0$. A reflected shock
propagates into the gas ($x>0$), leaving behind a flow of the
form~(\ref{special}) with $\rho_*(m, 0)=\rho_0(\gamma+1)[\Phi(m)\sqrt{\cos
m}+\Phi'(m)\sqrt{(\gamma-1)\cos m}]^{-2}$ and $A = \rho_0v_0^2(\gamma+1)/4$. The
function $\Phi(m)=\int_0^m\exp\left[-m'\sqrt{\gamma-1}\right]dm'/\cos^2m'$
determines the Lagrangian shock coordinate $m_s(t)$ implicitly by
$t(m_s)=2\Phi(m_s)/\left[\rho_0v_0(\gamma+1)\right]$. The collisional energy
loss slows down the gas reflection from the wall, producing a density blowup at
$m=0$ and $t=t_c=2\sqrt{\gamma-1}/\left[\rho_0v_0(\gamma+1)\right]$.

These exact solutions (note that they are not self-similar) describe a density
blowup $\sim (t_c-t)^{-2}$ at \textit{any} $\Lambda>0$. As $p$ is finite (and
non-zero), $T$ vanishes at the singularity. The gas velocity is finite, while
its $x$-derivative diverges as
$(t_c-t)^{-1}$. 
In the Eulerian coordinate $x$ the blowup of $\rho$ occurs on a shrinking
interval $\Delta(t) \sim(t_c-t)^{5/2}$, while $\rho_*\sim x^{-4/5}$ and does not
depend on time at $\Delta(t)\ll |x| \ll 1$. It is instructive to compare this
new singularity with the well-known \textit{free-flow} singularity
\cite{Whitham}, where the density blows up as $(t_c-t)^{-1}$, the shrinking
interval is $\Delta\sim (t_c-t)^{3/2}$, and the power law tail of the density at
singularity is $\sim x^{-2/3}$. For the free-flow singularity it is the
Lagrangian velocity, rather than the Lagrangian acceleration, that is constant
in time.

\begin{figure}%
\includegraphics[width=7.5 cm,clip=]{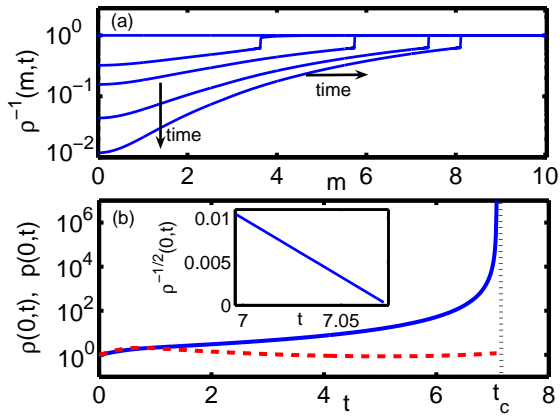}
\caption{The inverse density profiles at times $0$, $2.06$, $3.76$, $5.36$, and
$6.26$ (a) and the evolution of the density (the solid line) and pressure (the
dashed line) at the singularity point $m=0$ (b). The inset depicts
$\rho^{-1/2}(m=0, t)$ close to $t=t_c$. The parameters are $\gamma=2$, $M=1$ and
$\Lambda=0.5$. }\label{A2}
\end{figure}

\begin{figure}%
\includegraphics[width=5.5 cm,clip=]{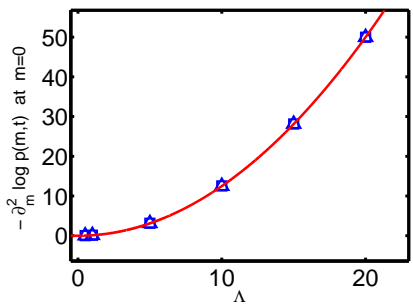}
\caption{
$-\partial_m^2\,\ln p(m,t)$ at $m=0$ and $t\simeq t_c$ vs. $\Lambda$ for $M=1$
(squares) and $M=10$ (triangles). The solid line shows $\Lambda^2/8$.}\label{A3}
\end{figure}

Our numerical simulations with Eqs.~(\ref{massmomcons}) and (\ref{temperature})
showed that, for generic initial conditions, a finite-time density blow up
always occurs \cite{finitedens}. Remarkably, the local flow structure near the
singularity always coincides with that exhibited by the special solutions
(\ref{special}). One series of simulations used the initial condition $\rho(m,
0)=\rho_0$, $T(m,0)=T_0$ and $v(m,0)=-v_0\tanh\left[m/(\rho_0 l)\right]$ at
different values of two governing scaled parameters: the Mach number
$M=v_0/T_0^{1/2}$ and the cooling coefficient rescaled by $\rho_0 l/M$ (and
denoted by $\Lambda$ with no ambiguity). The peak density $\rho(0, t)$ exhibits
a finite-time blow up, see Fig.~\ref{A2}, even at a relatively small
$\Lambda=0.5$, whereas $p$ remains finite.
Moreover, $\rho^{-1/2}(0,t)$ goes to zero linearly as $t \to t_c$. We observed
these properties to hold at all $\Lambda$ and $M$ for which a high density is
reached within a reasonable computation time (generally $t_c \to \infty$ as
$\Lambda\to 0$). Figure~\ref{A3} shows that $-\partial_m^2\,\ln p$ at $m=0$ and
$t \simeq t_c$ vs. $\Lambda$ is equal to $\Lambda^2/8$: the prediction of
Eq.~(\ref{special}) with the restored $\Lambda$-dependence at $\gamma=2$.
The same features (the density blowup and the universal flow structure near the
singularity) are also observed when starting from a small-amplitude sinusoidal
density or velocity perturbation around the homogeneous state. These results
leave little doubt that the asymptotic behavior of the hydrodynamic fields near
the singularity is described by Eqs.~(\ref{special}).

All the above implies that the non-linear evolution of the clustering
instability, as described by the ideal hydrodynamics of a dilute inelastic gas,
brings about an infinite density. In sharp contrast with the linear stage of
clustering instability \cite{Goldhirsch,McNamara}, the gas pressure develops a
local \textit{maximum} at the point of the maximum density, and becomes
time-independent in the Lagrangian frame. 
Obviously, to produce a density blowup, the flow must develop a region with
$\partial_x v<0$. A negative shear at $t=0$ is necessary and sufficient for a
blowup in the limit of $\Lambda\gg 1$, other parameters being fixed. This is
because the pressure drops almost instantaneously there, resulting in a free
flow and a density singularity development in the regions where $\partial_x v<0$
\cite{ELM,Whitham}. As the velocity profile steepens, however, the compressional
heating builds up and arrests the temperature drop, while the pressure again
becomes important. The flow, having already developed a high density peak,
rearranges, and the blowup develops according to the new (non-free-flow)
scenario. The rearrangement occurs within a time interval shrinking to zero as
$\Lambda \to \infty$. At the end of this interval the temperature dives to zero.
So, as $\Lambda \to\infty$, one has $t_c\to |\min
\partial_x v(x, 0)|^{-1}$, the blowup time of the free flow
\cite{Whitham}. This explains the results of Ref. \cite{ELM} in a new light.
Furthermore, if excluded volume effects are taken into account then, for
sufficiently large $\Lambda$, the singularity remains of the free-flow type all
the way until the close-packing density is reached \cite{MP}. Still later
dynamics is describable then by the Burgers equation \cite{MP,Ben-Naim1}.


Now we return to our special solutions (\ref{special}). Remarkably, the
dilute-gas equations~(\ref{massmomcons}) and (\ref{temperature}) allow for an
effective description of the states of inelastic gas with infinite density
spikes. 
As differential equations of gas dynamics is indeterminate on a spike, one needs
to return to the \textit{integral} form of the mass, momentum and energy
balance. Let us continue the solution (\ref{special}) beyond $t=t_c$. We will
describe this procedure for $m\geq0$ \cite{even}. Assume that $\rho_*(m, t=0)$
has a maximum at $m=0$ and is monotone decreasing at $m>0$, vanishing at
$m=\pi/2$. Then at $t\geq t_c$ there is a point $0\leq m_*(t)<\pi/2$, implicitly
defined by $1=t\sqrt{A\rho_*(m_*(t), 0)\cos m_*(t)}$, such that
$\rho_*\left[m_*(t), t\right]$, as prescribed by Eq.~(\ref{special}), is
infinite. Now, at $t<t_c$ we formally put $m_*(t)=0$ so $m_*(t)$ is continuous
at $t=t_c$. Here is the solution in Eulerian coordinates, $\hat{\rho} (x, t)$,
$\hat{p}(x,t)$ and $\hat{v}(x,t)$, which holds at any $t\!>\!0$:
\begin{equation}
\hat{\rho} (x, t)=2m_*(t)\delta(x)+\rho_*(x, t), \;\;\;\hat{p}(x,t)=p_*(x,t),
\label{hat}
\end{equation}
where $\rho_*(x, t)$ and $p_*(x, t)$ are given implicitly at $x\geq 0$ by the
first line of Eqs.~(\ref{special}) and $x(m, t)=\int_{m_*(t)}^m dm'/\rho_*(m',
t)$. The gas velocity is
\begin{equation}
\hat{v}\!=\!-2\!\int_{m_*(t)}^m\!\!\sqrt{\frac{A\cos m'}{\rho_*(m', 0)}}dm'+2A
t\left[\sin m\!-\!\sin m_*(t)\right] \label{vhat}
\end{equation}
with the same $x(m,t)$. We observe that, at $t>t_c$, the solution includes a
finite mass $2 m_*(t)$ concentrated at the origin. It is easy to see that
Eqs.~(\ref{hat}) and (\ref{vhat}) solve Eqs.~(\ref{eqs0}) in the $m-$coordinate
at $m>m_*(t)$ [but not at $0\leq m<m_*(t)$]. As a result, they solve
Eqs.~(\ref{massmomcons}) and (\ref{temperature}) in the $x-$coordinate at $x>0$.
It is left to verify that the integral form of Eqs.~(\ref{massmomcons}) and
(\ref{temperature}) holds at $x=0$. The mass flux $\rho v$, which is an odd
function of $x$, has a discontinuity at $x=0$ that produces the condition ${\dot
m_*}=-\lim_{x\to +0}\rho(x, t)v(x, t)=-\lim_{m\to +m_*(t)}\rho(m, t)v(m, t)$
verifiable using L'Hopital's rule. The integral forms of the momentum and energy
equations demand continuity of the momentum and energy fluxes at $x=0$ [for
example, $\lim_{\epsilon\to 0}\int_{-\epsilon}^{\epsilon} \rho^2
T^{3/2}dx=\lim_{\epsilon\to 0}\int_{-\epsilon}^{\epsilon} \rho p T^{1/2}dx=0$ by
virtue of $T(x=0,t)=0$ and finiteness of $p(x=0,t)$]. It can be checked that
Eqs.~(\ref{hat}) and (\ref{vhat}) obey these demands. As $t \to \infty$, all
the gas mass accumulates at $x=0$. 
The same
continuation procedure, applied to our time-dependent solution with the shock,
gives an example of a flow where a density spike and a shock are both present.

According to molecular dynamic (MD) simulations \cite{MP}, the density buildup
in the clusters is arrested only when the close-packing density is reached. An
additional exact solution of Eqs.~(\ref{massmomcons}) and (\ref{temperature})
establishes a simple relation between the density spikes and the close-packed
clusters. This solution involves a constant gas flux from infinity. In a moving
frame the problem is equivalent to that of a piston entering, with speed $v_0$,
into a cold gas at rest. The latter problem was previously addressed by
Goldshtein \textit{et al.} \cite{Goldshtein}. Using an empiric equation of state
where the pressure diverges at the close-packing density, they found that, at
long times, the flow has three regions. First, there is a close-packed cluster,
in contact with the piston; the cluster size grows linearly in time. It is
followed by a region with non-trivial density, temperature and velocity
profiles, separated from the gas at rest by a shock. The shock speed tends to
$v_0$ in the dilute limit \cite{Goldshtein}. Importantly, this flow is
describable by the dilute equations (\ref{massmomcons}) and (\ref{temperature}).
In the piston reference frame we obtain $\rho=M(t)\delta(x)+\rho_{st}(x)$,
$v=v_{st}(x)$ and $T=T_{st}(x)$, where $\rho_{st}(x)$, $v_{st}(x)$ and
$T_{st}(x)$ form \textit{unique} steady state solution of Eqs.
(\ref{massmomcons}) and (\ref{temperature}) obeying the boundary conditions
$\rho(x=+\infty,t)=\rho_0$, $v(x=+\infty,t)=-v_0<0$ and $T(x=+\infty,t)=0$. In
this description the close-packed cluster of Ref. \cite{Goldshtein} becomes a
single point $x=0$ which contributes the term $M(t)\delta(x)$ to the gas
density. The mass $M(t)$ is coming from infinity, ${\dot M}=\rho_0v_0$, and
stored at $x=0$. The shock is steady at $x=L>0$, while at $x>L$ the gas is at
rest. In the region $0<x<L$ the mass and momentum fluxes are constant in space,
so one can express $v_{st}(x)$ and $T_{st}(x)$ via $\rho_{st}(x)$. The latter is
given implicitly by
\begin{equation}\label{profile}
x=\frac{1}{\Lambda \rho_0}\left\{(\gamma+3)
   \arcsin\sqrt{u}-\frac{\sqrt{u} \,\left[\gamma+3-(\gamma+1)u\right]}
   {\sqrt{1-u}}\right\}, \nonumber
\end{equation}
where $u=\rho_0/\rho_{st}$, and the $\Lambda$-dependence is restored. The
piston-to-shock distance is
$$L=\frac{1}{\Lambda
\rho_0}\left[(\gamma+3) \arcsin\sqrt{\frac{\gamma-1}{\gamma+1}}- \sqrt{8
(\gamma-1)}\right]\,.
$$
This characteristic length scale behaves as $\Lambda^{-1}$, so it is very small
at $\Lambda \gg 1$. The same feature holds for all solutions that we have
presented here.

In summary, ideal hydrodynamic equations for a freely cooling dilute inelastic
gas describe the formation of density singularities in a finite time.  What
becomes of the singularities in microscopic theory? When the local gas density
approaches the close-packing density, the singularities must get regularized and
give way to (finite-density and finite-size) close-packed particle clusters
\cite{MP,viscosity}. One way to put the regularization into the continuum theory
is to adopt an equation of state that diverges at the close-packing density
\cite{Goldshtein}. An alternative, suggested in this Rapid Communication, does
not demand any regularization. It is similar in spirit to the ideal gas dynamic
treatment of shock fronts (which in fact have a finite width) as
discontinuities, that is \textit{point singularities} of the first derivatives
of the hydrodynamic fields \cite{Landau}. As we have shown here, this
alternative yields a powerful effective description in terms of smooth flow
regions separated by point singularities: either ``ordinary" shocks, or density
spikes (close-packed clusters). It will be interesting to apply this description
to \textit{driven} flows of granular gases where theory can be compared with
experiment. Finally, a direct comparison of our exact solutions with MD
simulations is presently under way.

We thank G. Falkovich and P.V. Sasorov for useful comments. This work was
supported by the Israel Science Foundation (grant No. 107/05) and by the
German-Israel Foundation for Scientific Research and Development (Grant
I-795-166.10/2003).


\begin{references}
\bibitem{viscosity} Numerical simulations suggest that
an account of viscosity and heat conduction may not arrest the density
singularities \cite{ELM}.
\bibitem{even} We assume
that $p(x,t)$ and
$\rho(x,t)$ are even functions of $x$, while $v(x,t)$ is odd so that
$v(x=0,t)=v(m=0,t)=0$.
\bibitem{finitedens} The numerical simulations had to be terminated
at a very high but finite density.
\end{references}

\end{document}